\pdfoutput=1
\documentclass[
 aps,
 prb,%
 amsmath,amssymb,%
 showkeys,%
 reprint,%
 superscriptaddress,
 floatfix,%
]{revtex4-2}

\usepackage[product-units = power, range-units = single]{siunitx}
\usepackage{graphicx}
\usepackage{graphicx}% Include figure files
\usepackage{verbatim}
\usepackage{float}
\usepackage{epstopdf}
\usepackage{dcolumn}% Align table columns on decimal point
\usepackage{bm}% bold math
\usepackage{appendix}
\usepackage{siunitx}
\usepackage{lineno}
\usepackage{textcomp}
\usepackage{xr}
\usepackage{filecontents}
\usepackage[capitalize]{cleveref}

\DeclareSIUnit\BohrMagneton{$\mu_mathrm{B}$}
\DeclareSIUnit\formulaunit{f.u.}
\DeclareSIUnit\atomicunit{a.u.}
\DeclareSIUnit\arbunit{arb.un.}
\DeclareSIUnit\torr{Torr}
\DeclareUnicodeCharacter{2212}{-}
\DeclareUnicodeCharacter{2264}{-}

\usepackage{natbib}

\begin{document}
\title{Spin polarization and magnetotransport properties of systematically disordered $\mathrm{Fe}_{60}\mathrm{Al}_{40}$ thin films}

\author{K. Borisov}
\email[]{borisovk@tcd.ie}
\affiliation{School of Physics, CRANN and AMBER, Trinity College, Dublin 2, Ireland}
\author{J. Ehrler}
\affiliation{Institute of Ion Beam Physics and Materials Research, Helmholtz-Zentrum Dresden-Rossendorf, Bautzner Landstrasse 400, 01328 Dresden, Germany}
\author{C. Fowley}
\affiliation{Institute of Ion Beam Physics and Materials Research, Helmholtz-Zentrum Dresden-Rossendorf, Bautzner Landstrasse 400, 01328 Dresden, Germany}
\author{B. Eggert}
\affiliation{Faculty of Physics and Center for Nanointegration Duisburg-Essen (CENIDE),
University of Duisburg-Essen, Lotharstr. 1, 47051 Duisburg, Germany}
\author{H. Wende}
\affiliation{Faculty of Physics and Center for Nanointegration Duisburg-Essen (CENIDE),
University of Duisburg-Essen, Lotharstr. 1, 47051 Duisburg, Germany}
\author{S. Cornelius}
\affiliation{Institute of Ion Beam Physics and Materials Research, Helmholtz-Zentrum Dresden-Rossendorf, Bautzner Landstrasse 400, 01328 Dresden, Germany}
\author{K. Potzger}
\affiliation{Institute of Ion Beam Physics and Materials Research, Helmholtz-Zentrum Dresden-Rossendorf, Bautzner Landstrasse 400, 01328 Dresden, Germany}
\author{J. Lindner}
\affiliation{Institute of Ion Beam Physics and Materials Research, Helmholtz-Zentrum Dresden-Rossendorf, Bautzner Landstrasse 400, 01328 Dresden, Germany}
\author{J. Fassbender}
\affiliation{Institute of Ion Beam Physics and Materials Research, Helmholtz-Zentrum Dresden-Rossendorf, Bautzner Landstrasse 400, 01328 Dresden, Germany}
\author{R. Bali}
\affiliation{Institute of Ion Beam Physics and Materials Research, Helmholtz-Zentrum Dresden-Rossendorf, Bautzner Landstrasse 400, 01328 Dresden, Germany}
\author{P. Stamenov}
\affiliation{School of Physics, CRANN and AMBER, Trinity College, Dublin 2, Ireland}

\date{\today}

\begin{abstract}
We investigate the evolution of spin polarization, spontaneous Hall angle (SHA), saturation magnetization and 
Curie temperature of $B2$-ordered Fe$_{60}$Al$_{40}$ thin films under varying antisite disorder, induced by 
Ne$^{+}$-ion irradiation. The spin polarization increases monotonically as a function of ion fluence. A relatively high 
polarization of \SI{46}{\percent} and the SHA of \SI{3.1}{\percent} are achieved on \SI{40}{\nano\meter} thick films 
irradiated with 2~$\cdot$~10$^{16}$ ions/\si{\centi\metre\squared} at \SI{30}{\kilo\electronvolt}. 
An interesting divergence in the trends of the magnetization and SHA is observed for low disorder concentrations.
The high spin polarization and its broad tunability range make ion-irradiated Fe$_{60}$Al$_{40}$ a promising material for application in spin electronic devices.
\end{abstract}

\keywords{Spin polarization, Iron-Aluminium, Spintronics, Anomalous Hall Effect, Topological Hall effect, Irradiation effects, Thin films, Magnetometry, Andreev reflection}%Use showkeys class option if keyword
                              %display desired

\maketitle

\section{Introduction}
\label{sec:intro}
	The performance of magnetic and electronic devices depends critically on the crystallography and chemical composition of the constituent materials. Ion-irradiation is capable of modifying the magnetic properties of ordered compositions. Magnetic anisotropy direction can change from perpendicular-to-plane to in-plane upon irradiation of Co/Pt multilayers \cite{Chappert1998}, while the opposite effect (of perpendicular magnetic anisotropy enhancement) has been shown in FePt(001)\cite{Ravelosona2000}. 
	One composition, which exhibits significant sensitivity of its magnetic properties on chemical order is Fe$_{1-x}$Al$_x$. The magnetic behaviour of Fe$_{1-x}$Al$_x$ can be counterintuitive, in the sense that the fully ordered $B$2 structure is paramagnetic (with rather low ordering temperature) while the $A$2 is ferromagnetic, well above room temperature. This can be ascribed to the changes in the local environment, as treated by the crystal field model\cite{Wertheim1964, Takahashi1986}, where the average number of Fe-Fe 
nearest neighbours determines the strength of the exchange interaction and the 
magnetic moment per Fe atom.  The phase region $x~<~0.2$ is 
ferromagnetic and the magnetization follows the linear relation 
$\overline{\mu}=2.2\mu_{\mathrm{B}}(1-x)$/Fe (where $\mu_{\mathrm{B}}$ is 
the Bohr magneton)\cite{Alcazar1987}. The magnetic 
behaviour in the intermediate concentration region $0.2~<~x~<~0.4$ depends on the preparation conditions and the degree of chemical order\cite{Correa2010, Taylor1958, Yelsukov1992, Sato1959, Vincze1971, Shull1976, Shiga1976}, and is attributed to the competition between Fe-Fe ferromagnetic exchange and Fe-Al-Fe antiferromagnetic superexchange interactions.

This article is focused on the Fe$_{60}$Al$_{40}$ alloy in thin film form, 
which is paramagnetic at room temperature and is stable across a broad temperature range\cite{Taylor1958}. The ideal 
stoichiometric Fe$_{60}$Al$_{40}$ composition must be paramagnetic, 
however, antisite defects 
lead to finite spontaneous magnetization at low temperature.
In thin films, chemical disorder produced by ion-irradiation of $B$2-ordered Fe$_{60}$Al$_{40}$, has been shown to generate  ferromagnetism via the transition from the $B$2 to the $A$2 structure\cite{Menendez2008,LaTorre2018, Ehrler2020}.
The ion energy can be selected such that it results in a homogeneous magnetization distribution in sufficiently thin films\cite{Roder2015}.
The $B$2 to $A$2 transition has been recently exploited for embedding single
nanoscale magnetic objects in a non-magnetic matrix\cite{Nord2019}, large-area
magnetic patterns\cite{Krupinski2019}, reversible magnetic
  writing \cite{Ehrler2018}, as well as in tuning the
magnetic anisotropy \cite{Schneider2019}.
 High-resolution, down to 
\SI{40}{\nano\metre}, magnetic stripe-patterning has been 
previously demonstrated, which opens the possibility of \textit{writing} 
planar GMR spin-valve structures \cite{Bali2014}. 
However, properties such as the Fermi level spin-polarization, $P(E_{\mathrm{F}})$, spontaneous Hall angle (SHA) and magnetoresistance (MR) are unknown. 

Here we investigate the variation of  $P(E_{\mathrm{F}})$, SHA and MR with increasing ion-fluence.
This manuscript aims to understand the scaling of relevant magneto-transport parameters with irradiation controlled disordering. 
The viability of controlling $P(E_{\mathrm{F}})$ by ion irradiation is examined. We analyze in detail the MR, the SHA, as well as the magnetization. The MR and SHA behaviour do not strictly follow the magnetization due to band structure modification with the lattice disordering. We show that ion-irradiation can be 
used to significantly enhance the spin polarization. These Fe$_{60}$Al$_{40}$ thin films can be, therefore, interesting for application in spin-detectors as they show large and tunable SHA.   

\section{Experimental methods}
The samples are grown by magnetron sputtering from a single Fe$_{60}$Al$_{40}$ target onto Si(001)/SiO$_{2}$ (\SI{150}{\nano\metre}) substrates at room temperature in an Ar-atmosphere of \SI{1e-3}{\milli\bbar} in a chamber with 
base pressure \SI{1e-8}{\milli\bbar}. The thickness of all the samples is 
close to \SI{40}{\nano\metre} and the surface of all of them is not protected by 
a capping layer as they are self-passivating. The as-deposited thin films are annealed at \SI{500}{\degreeCelsius} in a high vacuum furnace with base pressure \SI{5e-7}{\milli\bbar} for \SI{1}{\hour} in order to form the ordered $B$2 phase. Irradiation was performed with Ne$^{+}$ ions with energy 
\SI{30}{\kilo\electronvolt} at the Ion Beam Center (IBC), Helmholtz-Zentrum Dresden-Rossendorf (HZDR). The thicknesses of the films are measured by X-ray reflectivity (XRR).
All magnetotransport 
measurements are performed on blanket films in van der Pauw geometry in a 
Quantum Design Physical Property Measurement System (PPMS) with the 
standard resistivity option or with Keithley 2400 source-meters in the temperature range of \SI{10}{\kelvin} to \SI{300}{\kelvin} and applied field of up  to $\mu_{0}H~=~\pm\SI{14}{\tesla}$. 
The AHE and magnetoresistance (MR) are calculated
from the raw Hall effect data ($\rho_{xy}$) after field-antisymmetrization 
and field-symmetrization:
\begin{align*}
\mathrm{AHE}(\mu_{0}H) &= (\rho_{xy}(\mu_{0}H) - \rho_{xy}(-\mu_{0}H))/2, \\
\mathrm{MR}(\mu_{0}H) &= (\rho_{xy}(\mu_{0}H) + \rho_{xy}
(-\mu_{0}H))/2.
\end{align*}
The spontaneous Hall angle (SHA) is determined by normalizing the amplitude of the AHE 
with the longitudinal resistivity ($\rho_{xx}$).
Point Contact Andreev Reflection (PCAR)  lock-in-based differential conductance $G(V)$ spectroscopy is performed within a purpose-built electrical insert into the PPMS, at $T = \SI{2.0}{\kelvin}$. The analysis is done within the modified Blonder-Tinkham-Klapwijk model (mBTK)\cite{Strijkers2001, Blonder1982}. A comprehensive description of the data acqusition, processing and fitting routines is given elsewhere\cite{Strijkers2001, Stamenov2012, Borisov2016}.  
Each thin film is gently cleaned with low energy Ar plasma (\SI{300}{\watt\per\dm\squared}, incident at \SI{30}{\degree}  to the surface) before PCAR 
measurement, in order to remove the passivation surface oxide layer and then swiftly 
transferred to the measurement cryostat. Magnetometry is measured in a Quantum Design Magnetic 
Properties Measurement System (MPMS) at $T = \SI{4}{\kelvin}$ - \SI{300}{\kelvin} 
with the reciprocating sample option (RSO) and at higher temperatures from 
\SI{300}{\kelvin} to \SI{700}{\kelvin} with the oven insert. Further measurements are provided in the Supplementary Information.

\section{Experimental results}

\subsection{Magnetometry}
\begin{figure}[htb!]
\includegraphics[width=\columnwidth]{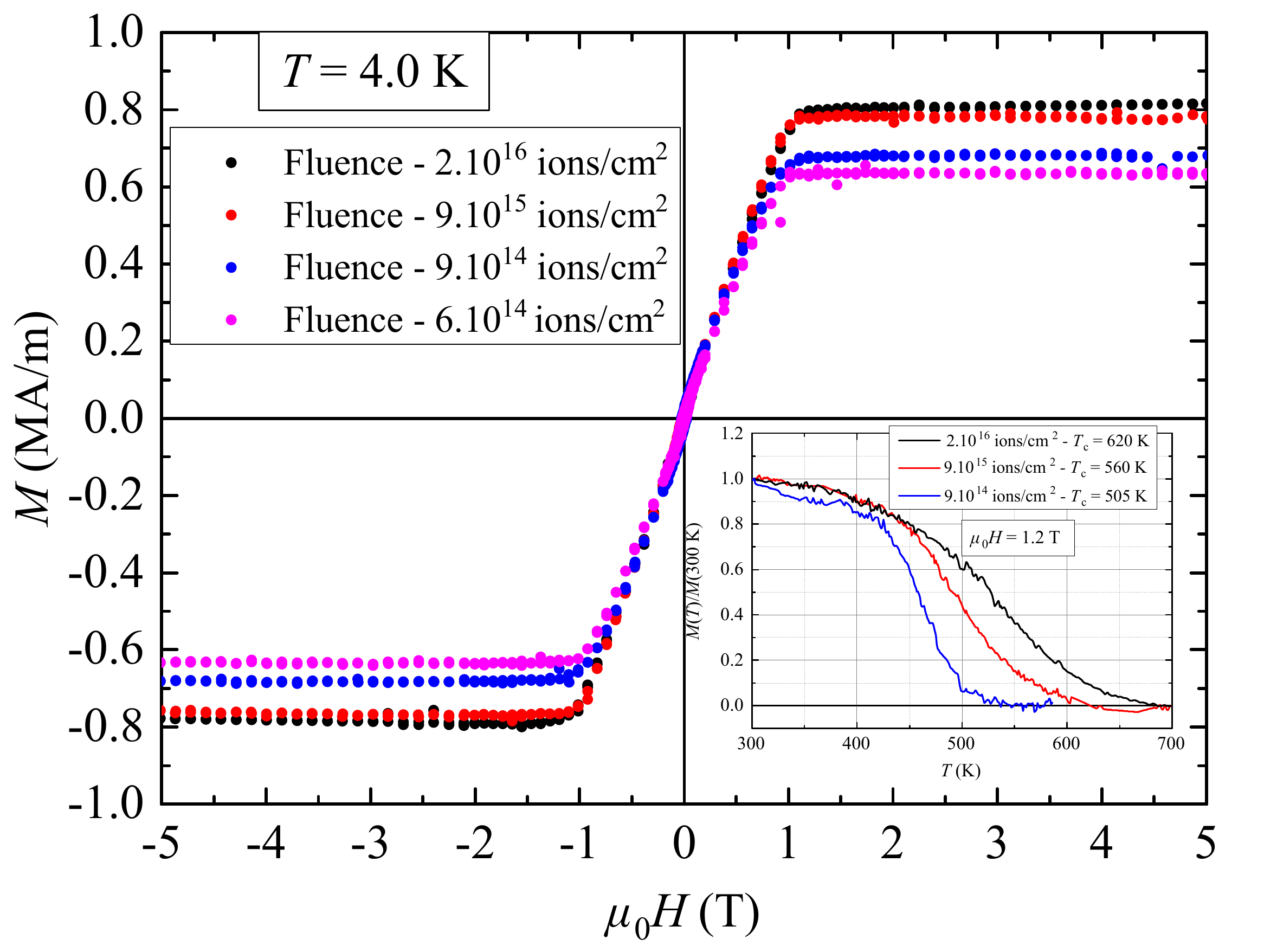}
\caption{ 
Hysteresis curves of four samples with different irradiation fluences, measured at \SI{4}{\kelvin}, with 
the magnetic field applied perpendicular to the plane. Inset: Curie temperature 
measurements of three samples with different irradiation fluences with  in-plane magnetic field $\mu_{0}H~=~\SI{1.2}{\tesla}$.}
\label{fig:SQUID_FeAl}
\end{figure}

First we demonstrate the changes of the saturation 
magnetization and the Curie temperature with increasing disorder (Fig.~\ref{fig:SQUID_FeAl}).
Thermal diffusion can result in reordering of the films during the $M(T)$ measurement, and for this reason these measurements were performed last.
The saturation magnetization demonstrates consistent increase as a function of the irradiation 
fluence: from \SI{20}{\kilo\ampere\per\metre} for the well-ordered $B_{\mathrm{2}}$ Fe$_{60}$Al$_{40}$ film\cite{Bali2014} to \SI{800}{\kilo\ampere\per\metre} (for fluence of 2~$\cdot$~10$^{16}$ ions/\si{\centi\metre\squared}). 
The magnetocrystalline anisotropy, $K_{1}$, is extracted from the apparent anisotropy field after demagnetizing field correction. This results in $K_{1}~=~\SI{50}{\kilo\joule\per\metre\cubed}$ (for 6~$\cdot$~10$^{14}$ ions/\si{\centi\metre\squared}) and 
$K_{1}~=~\SI{14}{\kilo\joule\per\metre\cubed}$ (for 2~$\cdot$~10$^{16}$ ions/\si{\centi\metre\squared}). 
The magnetocrystalline anisotropy is small (as expected for a cubic 
structure) and for the lowest fluence sample is close to the 
value of bulk $\alpha$Fe [110]\cite{Lawton1948}.
 
The Curie temperature is shown to increase with the irradiation fluence from $T_{\mathrm{C}} \approx \SI{505}{\kelvin}$ for 9 $\cdot$ 10$^{14}$ ions/\si{\centi\metre\squared} to 
\SI{620}{\kelvin} for 
2 $\cdot$ 10$^{16}$~ions/\si{\centi\metre\squared} (see inset of Fig.(\ref{fig:SQUID_FeAl})). There is a positive 
correlation between the irradiation fluence and the Curie temperature enhancement as is the case with the saturation magnetization. 

\subsection{Spin polarization measurements}

\begin{figure}[htb!]
\includegraphics[width=\columnwidth]{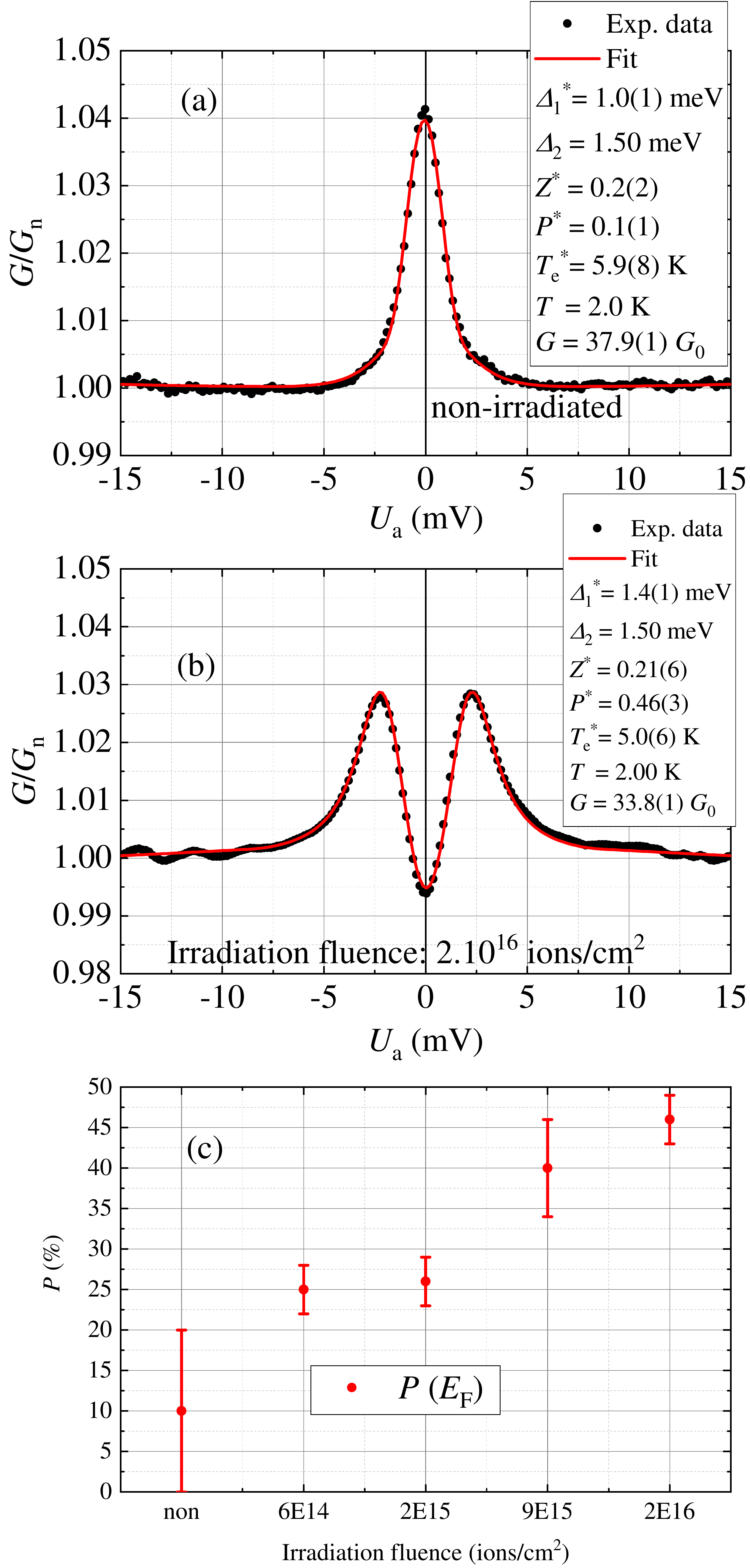}
\caption{PCAR spin polarization measurements of Fe$_{60}$Al$_{40}$ thin films with 
different irradiation fluences. The values are evaluated within the ballistic transport regime due to the high point contact resistance in these measurements\cite{Mazin2001, Gramich2012}. Panel (a): PCAR of an as-deposited, non-irradiated sample 
with small spin polarization of $P~\approx~\SI{10}{\percent}$. 
Panel (b): PCAR of a sample irradiated with the 
highest fluence of 2 $\cdot$ $10^{16}$ ions/\si{\centi\metre\squared} with moderately high spin polarization $P~\approx~\SI{46(3)}{\percent}$. A star superscript indicates fitted parameters. Panel (c): 
Spin polarization values as a function of the irradiation fluence.}
\label{fig:PCAR_FeAl}
\end{figure}

Fermi level spin polarization, $P(E_{\mathrm{F}})$, is investigated in samples with different irradiation fluences by the well-established PCAR technique\cite{Soulen1999}.
The spin polarization of five samples, the as-deposited composition and four with 
different irradiation fluences of up to  2 $\cdot$ 10$^{16}$ ions/\si{\centi\metre\squared}, is measured.
Representative PCAR spectra along with mBTK fits on the $B$2 ordered sample and the $A$2 ordered one with the highest irradiation fluence are plotted in Fig.~\ref{fig:PCAR_FeAl} (a) and (b), respectively, where the extracted parameters from the analysis are superconducting proximity gap ($\Delta_{1}^{*}$), barrier strength ($Z^{*}$), spin polarization ($P^{*}$), and effective electron temperature ($T_{\mathrm{e}}^{*}$). The 
$B$2-ordered sample demonstrates an increase in the so-called zero bias anomaly, 
which is indicative of a very low spin polarization\cite{Soulen1999}.
The extracted value of 
$P \approx \SI{10}{\percent}$ with a significant error (\SI{10}{\percent}) in the 
fit demonstrates that the non-irradiated sample has no appreciable 
spin polarization at $E_{\mathrm{F}}$. PCAR has been demonstrated to not
be highly sensitive towards very low spin polarization values (at low superconducting proximity) and this explains 
the significant experimental error in this particular measurement\cite{Bugoslavsky2005}. 
Furthermore, four samples with increasing 
irradiation fluences are investigated. The spin polarization has been extracted 
to be \SI{25(3)}{\percent}, \SI{26(3)}{\percent}, \SI{40(6)}{\percent}, and 
\SI{46(3)}{\percent} for irradiation fluences of 
6 $\cdot$ $10^{14}$ ions/\si{\centi\metre\squared}, 
2 $\cdot$ $10^{15}$ ions/\si{\centi\metre\squared}, 
9 $\cdot$ $10^{15}$ ions/\si{\centi\metre\squared}, 
and 2 $\cdot$ $10^{16}$ ions/\si{\centi\metre\squared}, respectively. The trend 
towards higher spin polarization shows a positive correlation 
between the irradiation fluence and the difference between the spin-up and spin-down 
density of states at the Fermi level. 
Further insight into the properties of the band structure at the Fermi level is given by the magnetotransport behaviour of the irradiated films.

\begin{figure*}[htb!]
\includegraphics[width=\textwidth]{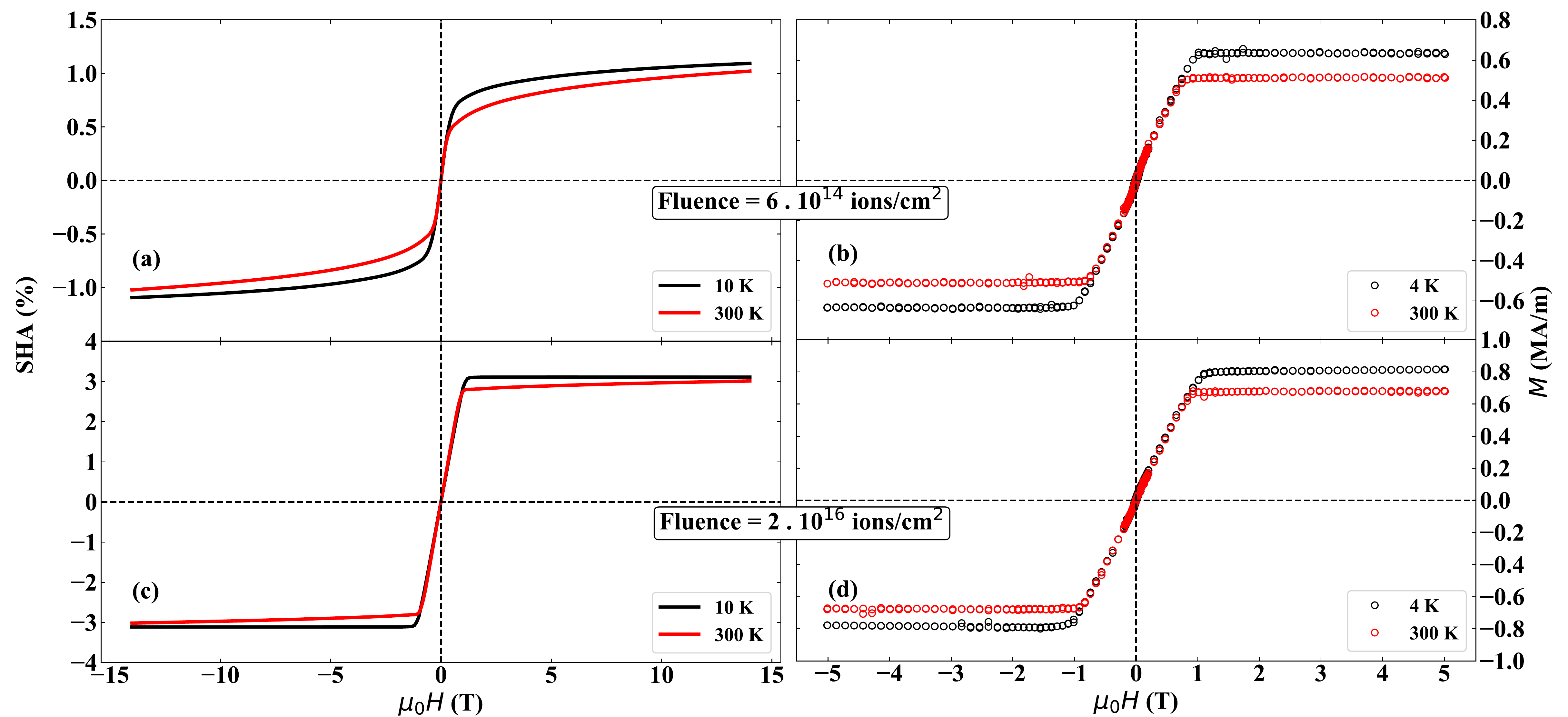}
\caption{Comparison between the SHA and SQUID magnetometry on a sample with low 
irradiation fluence (6 $\cdot$ $10^{14}$ ions/\si{\centi\metre\squared}) and high 
irradiation fluence (2 $\cdot$ $10^{16}$ ions/\si{\centi\metre\squared}). The 
magnetic field is always applied perpendicular to the sample plane. 
Panel~(a): SHA measurement at \SI{10}{\kelvin} and 
\SI{300}{\kelvin} for the low irradiation fluence sample. Panel~(b):~magnetization data at \SI{4}{\kelvin} and 
\SI{300}{\kelvin} for the same sample.
Panel~(c):~SHA 
measurement at \SI{10}{\kelvin} and \SI{300}{\kelvin} for the high irradiation 
fluence sample. Panel~(d):~magnetization data at \SI{4}{\kelvin} and \SI{300}{\kelvin} for the same sample. }
\label{fig:FeAl_4_panel}
\end{figure*}
\subsection{Non-saturating Spontaneous Hall Angle}

\begin{figure*}[htb!]
\includegraphics[width=\textwidth]{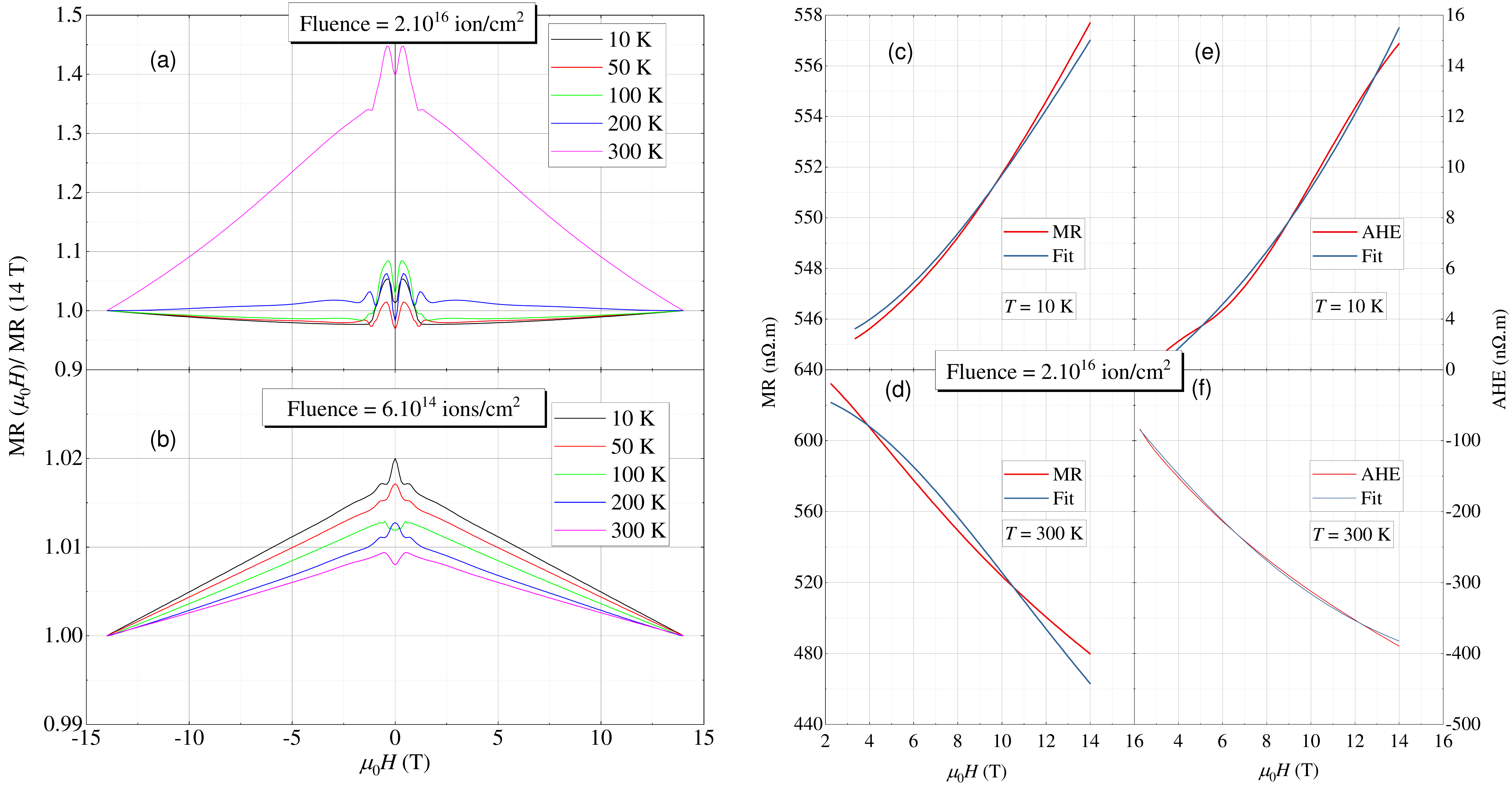}
\caption{Two carrier type analysis of magnetoresistance and AHE of the highest irradiation fluence sample (2 $\cdot$ 10$^{16}$ ions/\si{\centi\metre\squared}). Panel (a): MR of sample with irradiation fluence 2 $\cdot$ 10$^{16}$ ions/\si{\centi\metre\squared} at various temperatures. Panel (b): MR of sample with irradiation fluence 6 $\cdot$ 10$^{14}$ ions/\si{\centi\metre\squared} at various temperatures. The V-shaped feature at low field is an antisymmetrization artefact. Panel (c), (d): MR of the high fluence samples at \SI{10}{\kelvin} and \SI{300}{\kelvin} along with a two type carrier model fits. Panel (e), (f): AHE of the high fluence samples at \SI{10}{\kelvin} and \SI{300}{\kelvin} along with a two type carrier model fits. 
}
\label{fig:FeAl_6_panel}
\end{figure*}

We have measured significant disagreement between the SHA and the magnetization of the irradiated 
Fe$_{60}$Al$_{40}$ samples\cite{Nagaosa2010}.
The latter is most striking for the sample with the low irradiation fluence of 
6 $\cdot$ $10^{14}$ ions/\si{\centi\metre\squared} (see Fig.~\ref{fig:FeAl_4_panel} (a) and (b)).
Both the low and room temperature SHA do not reach 
saturation in \SI{14}{\tesla} in contrast to the SQUID magnetization for 
this sample. 
For the highest irradiation fluence, SHA is essentially saturated beyond $\mu_{0}H_{\mathrm{an}}~\approx~\SI{1.1}{\tesla}$ (Fig.~\ref{fig:FeAl_4_panel} (c)) as 
is the bulk magnetization (Fig.\ref{fig:FeAl_4_panel}(d)) at \SI{10}{\kelvin}. 
%The SHA reaches very high value of \SI{3.1}{\percent} in the greatest fluence sample. 
Interestingly, there is a pronounced discrepancy between the SHA and 
SQUID magnetometry beyond $\mu_{0}H_{\mathrm{an}}$ at \SI{300}{\kelvin}. 
The SHA exhibits an inflection point at $\mu_{0}H_{\mathrm{an}}$, 
however, the high field signal does not reach saturation even in 
\SI{14}{\tesla} (see red curve in Fig.~\ref{fig:FeAl_4_panel} (c)). 
The room temperature SHA has clearly a non-linear character and, hence, is attributed 
to anomalous Hall effect and not to ordinary Hall effect, 
\textit{i.e.} change in the carrier density between \SI{10}{\kelvin} and 
\SI{300}{\kelvin} (the ordinary Hall effect must be linear). 
The change in the high-field background beyond 
\SI{1}{\tesla}~($\mu_{0}H_{\mathrm{an}}$) evolves smoothly from a flat 
line (highest fluence) to a quasi-quadratic, non-saturating background 
(lowest fluence).

\subsection{Magnetoresistance}
The magnetoresistance (MR) of the samples with the highest and the lowest 
irradiation fluences at various temperatures is presented in 
Fig.~\ref{fig:FeAl_6_panel} (a) and (b). 
The MR of the highest irradiation fluence sample changes from negative to 
positive between \SI{10}{\kelvin} and \SI{300}{\kelvin}, which can indicate 
change of the dominant carrier type (for the most mobile bands). 
Pronounced increase in MR 
amplitude is seen at \SI{300}{\kelvin} which coincides with the non-
saturating Hall effect at the same temperature (see Fig~\ref{fig:FeAl_4_panel} (c)). 
On the other hand, the 6 $\cdot$ $10^{14}$ ions/\si{\centi\metre\squared} irradiation fluence sample exhibits constant 
positive magnetoresistance at all measured temperatures and, hence,
confirms a constant carrier type which is temperature 
independent (Fig.~\ref{fig:FeAl_6_panel} (b)).
The MR and Hall effect of the highest fluence sample are fitted within a two-carrier type model at \SI{10}{\kelvin} and 
\SI{300}{\kelvin}\cite{Alaria2010, Borisov2014} in 
Fig.~\ref{fig:FeAl_6_panel} (c) - (f). The model assumes multiple carriers types with effective masses equal to the free electron 
mass, therefore, negative $n_i$ means effective mass 
with opposite sign($i~=~2$) - a hole.
At low temperature, the first carrier type has
concentration $n_{1}(\SI{10}{\kelvin})~=~\SI{2.328e27}{\per\meter\cubed}$ and scattering 
time $\tau_{1}(\SI{10}{\kelvin})~=~\SI{2.24e-13}{\second}$, while the second carrier type 
has concentration $n_{2}(\SI{10}{\kelvin})~=~\SI{-2.213e29}{\per\meter\cubed}$ and 
scattering time $\tau_{2}(\SI{10}{\kelvin})~=~\SI{2.71e-14}{\second}$ . On the other hand 
at room temperature, the first carrier type has $n_{1}(\SI{300}{\kelvin})~=~\SI{1.37e29}{\per\meter\cubed}$ and $\tau_{1}(\SI{300}{\kelvin})~=~\SI{7.28e-14}{\second}$, while the second carrier type has 
$n_{2}(\SI{300}{\kelvin})~=~\SI{-1.54e28}{\per\meter\cubed}$ and 
$\tau_{2}(\SI{300}{\kelvin})~=~\SI{2.81e-13}{\second}$. The extracted
higher concentration of holes at low temperature 
confirms the already anticipated result from the negative MR. 
We can estimate the conductivity contributions, $\sigma_{i}~=~n_{i}e\mu_{i}$ with the mobility $\mu_{i}~=~e\tau_{i}/m_{\mathrm{e}}$, where $e$, $m_{\mathrm{e}}$ are electron's charge and mass. 
%\footnote{The effective mass of the carriers is essentially absorbed in the scattering times in the model.} 
It is evident that $\vert\sigma_{1}^{\mathrm{e}}(\SI{10}{\kelvin})\vert<\vert\sigma_{2}^{\mathrm{h}}(\SI{10}{\kelvin})\vert$ which confirms dominant conductivity contribution of holes at low temperature. Respectively, $\vert\sigma_{1}^{\mathrm{e}}(\SI{300}{\kelvin})\vert>\vert\sigma_{2}^{\mathrm{h}}(\SI{300}{\kelvin})\vert$ which testifies for dominant conductivity contribution of electrons at room temperature.
Analogous carrier type cross-over was recently reported for other transition metal alloys\cite{Gomes2019}.
The constancy of the dominating carrier type for the low irradiation 
fluence sample is corroborated within the two-type carrier analysis. No other sample exhibited MR sign change for 
$T~=~\SI{10}{\kelvin}-\SI{300}{\kelvin}$ apart from the maximum irradiation fluence one.

\subsection{Temperature Dependence of Spontaneous Hall Angle}
\begin{figure}[htb!]
\includegraphics[width=\columnwidth]{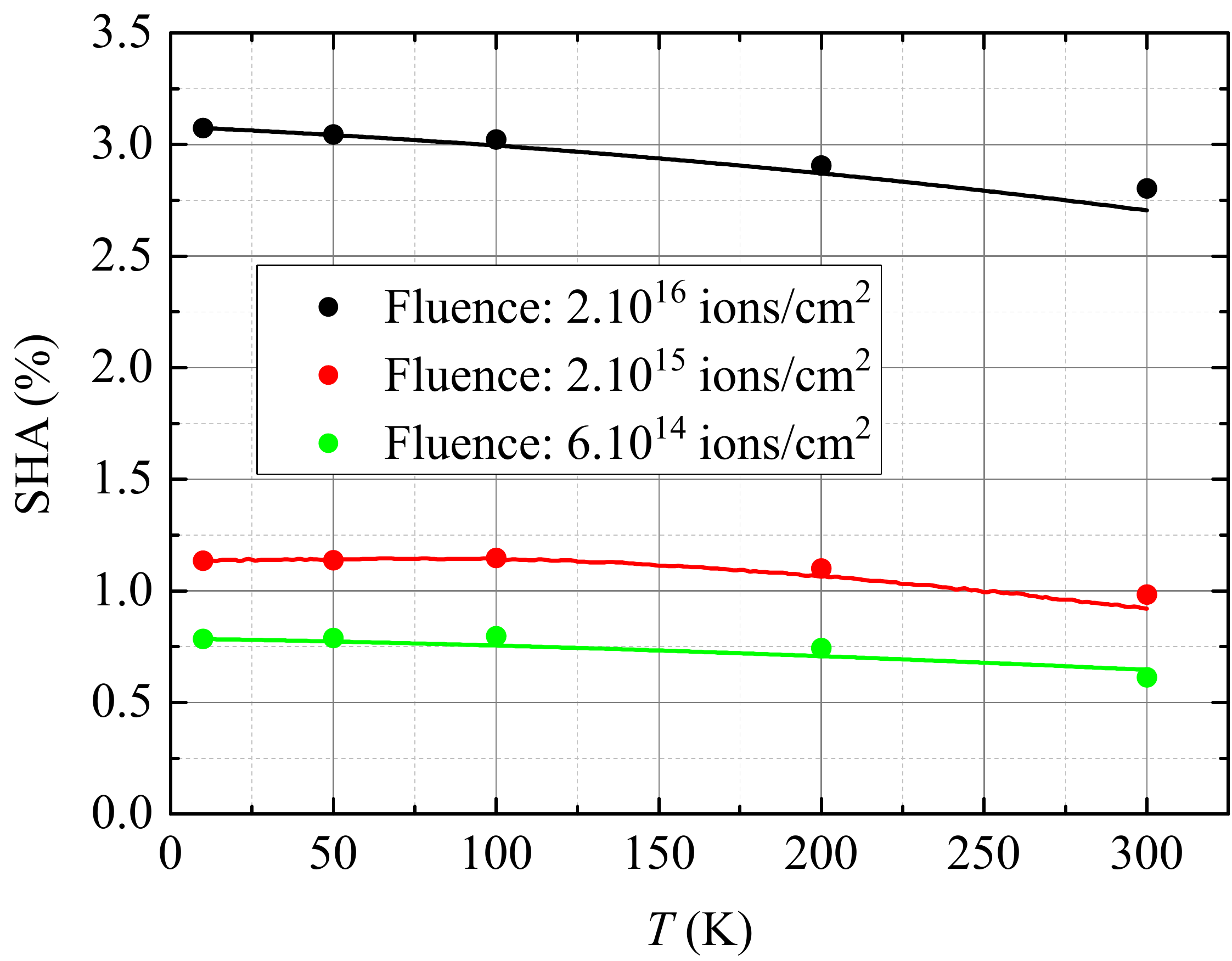}
\caption{Comparison between the magnetization and SHA temperature 
dependence for three samples with different irradiation fluences. The filled
dots represent the SHA. The solid lines are the 
magnetization data which have been normalized to the SHA(\SI{10}{\kelvin}) for each sample. The
magnetization and SHA signals are taken at \SI{1.2}{\tesla} for both panels. See Fig.~\ref{supp-fig:SHA_SQUID_Suppl} for further details.}
\label{fig:FeAl_SHA}
\end{figure}

The comparison between the normalized magnetization and the normalized SHA for three irradiation fluences is 
presented in Fig.\ref{fig:FeAl_SHA}. 
The irradiation dependence shows that SHA equals \SI{0.7}{\percent} for 6~$\cdot$~10$^{14}$ ions/\si{\centi\metre\squared}, increases to \SI{1.1}{\percent} for 2~$\cdot$~$10^{15}$ ions/\si{\centi\metre\squared} and reaches \SI{3.1}{\percent} for the sample irradiated with 2~$\cdot$~$10^{16}$ ions/\si{\centi\metre\squared}. 
Similar SHA values have been recently reproduced by first-principle calculation on the disorder effect in Fe$_{60}$Al$_{40}$\cite{Kudrnovsky2020}.
High SHA values, within the same order of magnitude, 
have been reported in amorphous Fe$_{0.79}$Gd$_{0.21}$ (\SI{5.8}{\percent}) 
\cite{McGuire1977}, amorphous Co-Tb (\SI{3.2}{\percent}) \cite{Kim2000}, 
$L$1$_{0}$-FePt (\SI{3.3}{\percent}) \cite{Yu2000}, Mn-Ga (\SI{5.7}{\percent}) 
\cite{Wu2010}, Mn$_2$Ru$_x$Ga (\SI{7.7}{\percent}) \cite{Thiyagarajah2015}, and 
amorphous Co$_{40}$Fe$_{40}$B$_{20}$ (\SI{2.3}{\percent}) \cite{Su2014}. 

\subsection{Discussion}
The magnetometry data in Fig.~\ref{fig:SQUID_FeAl} demonstrates consistent increase in the saturation magnetization as a function of the irradiation fluence, which is consistent with increasing antisite disorder. The saturation magnetization reaches \SI{800}{\kilo\ampere\per\metre} for the highest irradiation fluence of 2~$\cdot$~10$^{16}$ ions/\si{\centi\metre\squared}. In accordance with $M_{\mathrm{s}}$, the Curie temperature is determined to increase as well, up to \SI{620}{\kelvin} for the highest irradiation fluence.
Structural reordering towards $B$2 at such elevated temperatures have been reported\cite{Ehrler2019}. 
Note that thermal reordering is known to commence at $\approx~\SI{400}{\kelvin}$\cite{Ehrler2019}. The re-ordering occurs simultaneously with the Curie transition, and further lowers the magnetization.

The PCAR measurements (see Fig.~\ref{fig:PCAR_FeAl}) demonstrate a pronounced enhancement of the Fermi level spin polarization from very low (\SI{10}{\percent}) for the non-irradiated sample to respectably high value ( \SI{46}{\percent}) for the highest irradiation fluence of 2 $\cdot$ $10^{16}$ ions/\si{\centi\metre\squared}.
Higher spin polarization
implies stronger $s-d$ exchange interaction, which is in line with the higher Curie temperatures of the more irradiated compositions (see Fig.~\ref{fig:SQUID_FeAl} (b)).
The measured Fermi level spin polarization of up to 
\SI{46(3)}{\percent} is similar to the value of pure Fe \cite{Karthik2009} and 
close to the value of the exemplary high-moment Co-Fe compositions \cite{Parkin2004, Karthik2009}. 
The comparison between the SQUID magnetometry and the AHE measurements evidenced in Fig.~(\ref{fig:FeAl_4_panel}) shows intriguing disagreement.
A viable explanation is that the splitting of the $s$-bands is influenced by the irradiation fluence and temperature. The conduction process is dominated by the more mobile, $s$-electrons with the accompanying spin splitting of 
the $s$ density of states, while the bulk magnetization is determined by 
the splitting of the $d$ states. The latter implies
essentially that the $s$-$d$ coupling, through which the conduction electrons 
become spin polarized, is smaller for the samples irradiated with lower fluence.
Hence, the non-saturating AHE contribution is 
attributed to non-polarized (random spin orientation), Al-related $s$-bands. For 
low irradiation fluences, these bands have high contribution to the Fermi 
level transport. The Fe $s$-bands are not sufficiently split in this case (compared to those in $\alpha$Fe) 
and they are responsible for the AHE signal up to 
$\mu_{0}H_{\mathrm{an}}$. For $\mu_{0}H~>~\mu_{0}H_{\mathrm{an}}$, the magnetic field acts on the Al $s$-electrons, their spins are gradually more aligned with the sample magnetization direction and this leads to the AHE increasing in magnitude up to and above \SI{14}{\tesla}. 
For higher irradiation fluences, the two Fe $s$-bands are split stronger and one of them dominates the Fermi level transport. This makes the Al, $s$-bands contribution small and can explain why the AHE behaviour follows closely the magnetization (low temperature in Fig.~\ref{fig:FeAl_4_panel} (c) and (d)). For the highest fluence at room temperature, thermal activation reduces the effective Fe, $s$-band splitting (\textit{i.e.} spin polarization) and this allows the Al, $s$-bands to exhibit their small, non-saturating contribution (\SI{300}{\kelvin} in 
Fig.~\ref{fig:FeAl_4_panel} (c)).
Therefore, our interpretation is that the irradiation modifies the effective $s$-$d$ coupling strength and governs the Fe $s$-bands splitting for this 
composition. This conjecture is in line with the PCAR data, presented in 
Fig.~\ref{fig:PCAR_FeAl}, which demonstrates that the lower fluence samples have lower spin polarization, therefore, smaller spin-splitting for the conduction bands. Further resistivity measurements are provided in Fig.~\ref{supp-fig:FeAl_Rxx}.

The temperature evolution of the magnetoresistance supports this argument (Fig.~ \ref{fig:FeAl_6_panel}).
For the highest irradiation fluence (2 $\cdot$ $10^{16}$ ions/\si{\centi\metre\squared}), the MR changes sign at $T~\approx~\SI{150}{\kelvin}~$ ($\Delta E~\approx~\SI{15}{\milli\electronvolt}$). This observation is attributed to temperature dependent Fermi level, which changes its position between two distinct density of states mini-pockets with energy spacing of the order of $\Delta E$ (see Fig.~ \ref{fig:FeAl_6_panel}(a)). The latter is confirmed within a two-carrier model fitting of the MR and the AHE (Fig.~ \ref{fig:FeAl_6_panel} (c)-(f)). The low temperature MR sign change from positive to negative, as the number of Fe-Fe nearest neighbours increases due to irradiation, is reproduced by electronic structure calculations\cite{Kudrnovsky2020}.

The comparison 
between the normalized magnetizations and the SHAs temperature decrease shows similar behaviour and similar small reduction for all samples from 
\SI{10}{\kelvin} to 
\SI{300}{\kelvin}\cite{Ye2012} (Fig.~\ref{fig:FeAl_SHA}). 
The latter is a strong point that magnetotransport properties vary weakly with temperature which is an important material prerequisite for integration in spin electronic devices.
Most prominent is the SHA temperature decrease for the highest fluence which is significantly smaller than the magnetization reduction (see black line in Fig.~\ref{fig:FeAl_SHA}).
This indicates that the spin polarization depends on a complicated 
spin-split density of states structure close to $E_{\mathrm{F}}$. 
This observation may open new pathways for engineering spin-transport. The Fermi level spin polarization $P(E_{\mathrm{F}})$ may increase further in this $A$2-ordered alloy despite the fact that the value of Fe is already reached, because it is the DOS contributions of both Fe and Al bands and their hybridization, which determine $P(E_{\mathrm{F}})$ in Fe$_{60}$Al$_{40}$, and not purely the $s$-$d$ exchange interactions. 

\section{Conclusion}
Spin polarization, spontaneous Hall angle, saturation magnetization and Curie 
temperature are shown to increase in Fe$_{60}$Al$_{40}$ thin films upon $B$2-$A$2 crystal structure transition 
induced by irradiation with Ne$^{+}$ ions. For irradiation fluence of 
2 $\cdot$ $10^{16}$ ions/\si{\centi\metre\squared}, the spin polarization reaches 
\SI{46(3)}{\percent}, the spontaneous Hall angle - \SI{3.1}{\percent}, and the Curie 
temperature - \SI{620}{\kelvin}. 
Ion irradiation is demonstrated as an efficient method to increase the spin polarization and to modify the $s$-$d$ splitting in thin 
Fe$_{60}$Al$_{40}$ films. The discrepancy between the high-field non-saturating anomalous Hall effect and the magnetization is attributed to high electrical transport contribution from non-polarized, aluminium-related bands for low irradiation fluences. Such AHE-magnetization disagreement could be taken as an indication for lower spin polarization.
Even though it is possible to account for some of the observed discrepancy between the magnetization and the AHE as being due to the suppression of the topological Hall effect present in the $B$2 ordered phase, the level of difference in electron-phonon scattering observed between the highest and lowest irradiation fluences, for which the room temperature saturation magnetization is properly developed, does not seem to corroborate the same as the dominant contribution. In view of this, it seems likely, that in other intermetallic systems with weak anisotropy and large saturation moment, the reported differences between apparent approach to saturation, as seen in magnetization and Hall effect, may have other explanations, but the topological Hall effect.

The measured spin polarization 
of \SI{46}{\percent} implies that such irradiated compositions can be used as 
spin injectors/detectors in GMR spin-valves and novel magnetic circuit designs.
Spin diffusion length measurements of a non-irradiated sample and the energy dependence of spin-split density of states of irradiated samples could be interesting future research directions, which will help optimize the irradiation and annealing strategies.
The irradiation-induced 
ferromagnetic state in Fe$_{60}$Al$_{40}$ with the accompanying 
beneficial magnetotransport properties make this composition potential 
for integration in novel magnetic device geometries with high design 
flexibility and reduced processing complexity.  

\begin{acknowledgments}
K. B. and P. S. would like to acknowledge financial support from Science Foundation Ireland within SSPP (11/SIRG/
I2130), NISE (10/IN1/I3002) and AMBER programme. 
RB and HW acknowledge funding by the Deutsche Forschungsgemeinschaft (DFG) -
322462997 (BA 5656/1-2 | WE 2623/14-2).
K.B., C.F. and P.S. acknowledge support within European Union
Horizon 2020 research and innovation programme under
Grant Agreement No. DLV-737038 (TRANSPIRE). 
Support by the Ion Beam Center (IBC) at HZDR is gratefully acknowledged.
We are thankful to Prof. J.M.D. Coey, Dr. M. Venkatesan, Dr. P. Tozman and Dr. S. Khmelevskyi for fruitful discussions. 
\end{acknowledgments}

\bibliography{FeAl_PRB}

\makeatletter\@input{FeAl_suppl_aux.tex}\makeatother
\end{document}

\title{Supplementary Information for\\ ``Spin polarization and magnetotransport properties of systematically disordered $\mathrm{Fe}_{60}\mathrm{Al}_{40}$ thin films''}

\author{K. Borisov}
\email[]{borisovk@tcd.ie}
\affiliation{School of Physics, CRANN and AMBER, Trinity College, Dublin 2, Ireland}
\author{J. Ehrler}
\affiliation{Helmholtz-Zentrum Dresden-Rossendorf, Institute of Ion Beam Physics and Materials Research, Dresden, Germany}
\author{C. Fowley}
\affiliation{Helmholtz-Zentrum Dresden-Rossendorf, Institute of Ion Beam Physics and Materials Research, Dresden, Germany}
\author{B. Eggert}
\affiliation{Faculty of Physics and Center for Nanointegration Duisburg-Essen (CENIDE),
University of Duisburg-Essen, Lotharstr. 1, 47051 Duisburg, Germany}
\author{H. Wende}
\affiliation{Faculty of Physics and Center for Nanointegration Duisburg-Essen (CENIDE),
University of Duisburg-Essen, Lotharstr. 1, 47051 Duisburg, Germany}
\author{S. Cornelius}
\affiliation{Helmholtz-Zentrum Dresden-Rossendorf, Institute of Ion Beam Physics and Materials Research, Dresden, Germany}
\author{K. Potzger}
\affiliation{Helmholtz-Zentrum Dresden-Rossendorf, Institute of Ion Beam Physics and Materials Research, Dresden, Germany}
\author{J. Lindner}
\affiliation{Helmholtz-Zentrum Dresden-Rossendorf, Institute of Ion Beam Physics and Materials Research, Dresden, Germany}
\author{J. Fassbender}
\affiliation{Helmholtz-Zentrum Dresden-Rossendorf, Institute of Ion Beam Physics and Materials Research, Dresden, Germany}
\author{R. Bali}
\affiliation{Helmholtz-Zentrum Dresden-Rossendorf, Institute of Ion Beam Physics and Materials Research, Dresden, Germany}
\author{P. Stamenov}
\affiliation{School of Physics, CRANN and AMBER, Trinity College, Dublin 2, Ireland}
\date{\today}

\date{\today}
\maketitle

\section{Electrical transport properties}
\label{sec:Rxx}

\begin{figure}[htb!]
\includegraphics[width=\columnwidth]{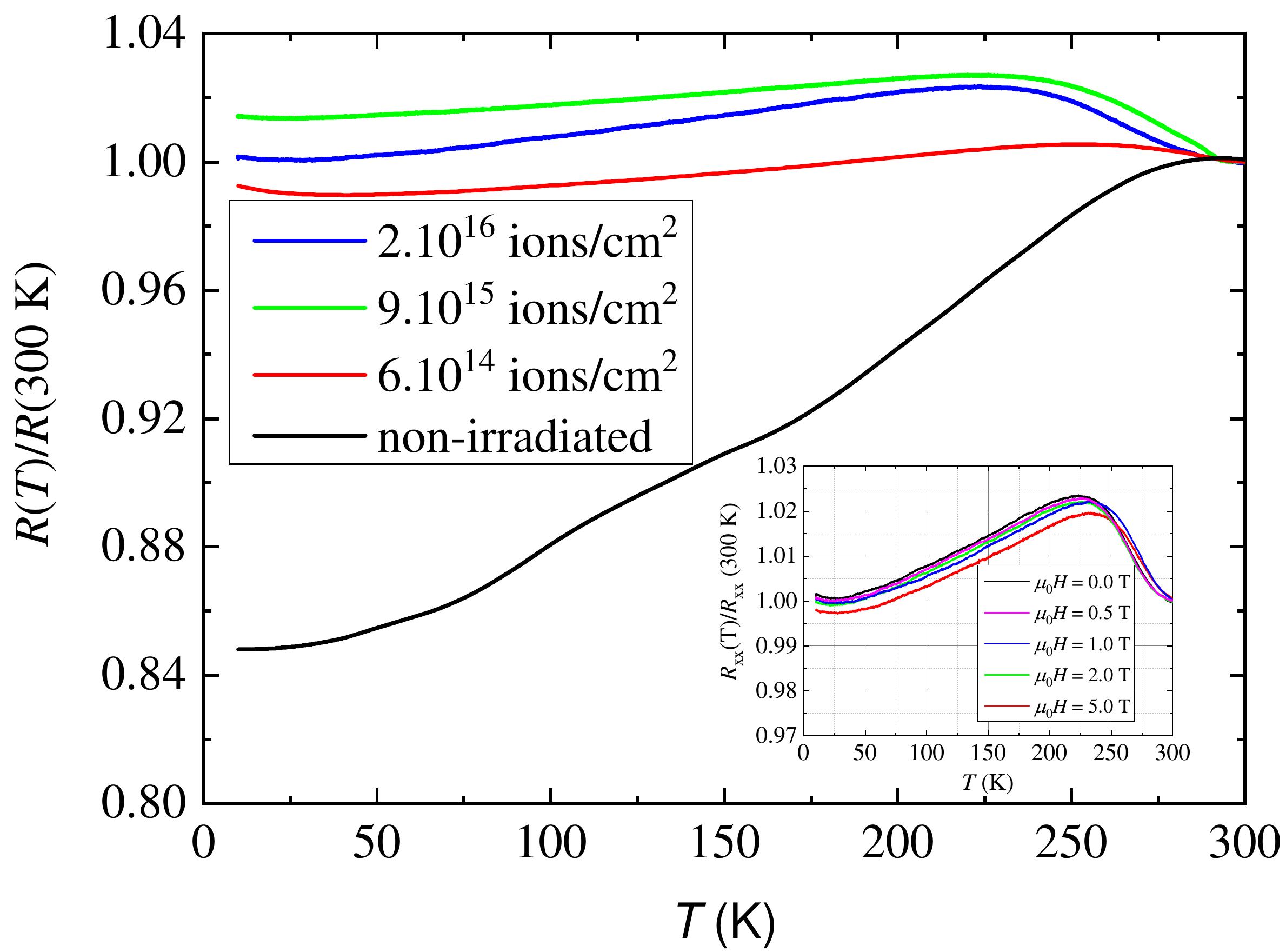}
\caption{Normalized resistance temperature dependence for samples with 
four different fluences in zero magnetic field. 
Inset: Normalized resistance temperature dependence in different 
applied magnetic field perpendicular to the surface of the sample irradiated with 2~$\cdot$~$10^{16}$~ions/\si{\centi\metre\squared}. The small differences between the curves is due to the inevitable AHE contribution during a longitudinal resistance measurement in van der Pauw configuration. 
}
\label{fig:FeAl_Rxx}
\end{figure}

The normalized longitudinal resistance-temperature evolution (NR~$(T)~=~R~(T)/R~(\SI{300}{\kelvin})$) of four samples with different irradiation fluences is presented in Fig.~\ref{fig:FeAl_Rxx}. The non-irradiated sample 
demonstrates the expected resistance decrease on cool down (with NR~$(\SI{10}{\kelvin})~\approx~0.85$, RRR~$\approx$~1.2) 
typical for high quality sputter-deposited metallic films\cite{Bainsla2014, Betto2017}, where the electron-phonon scattering dominates. The irradiation fluence of 6~$\cdot$~$10^{14}$~ions/\si{\centi\metre\squared} results in higher NR~$(\SI{10}{\kelvin})$ of $0.96$, as well as 
small resistance upturn at low temperatures. 
The irradiation induces a $B$2 to $A$2 crystal structure transition and higher average number of Fe-Fe nearest neighbours. An increased NR is a typical indication of that\cite{Mooij1973}.
%This sample does not exhibit a pronounced metallic resistance decrease on cool down, which is due to the crystallographic $B$2-$A$2 transition induced by the ion irradiation. 
The phonon 
scattering process has a decreased contribution in this case because the electron's mean free path is reduced\cite{Gurvitch1981} while the crystallinity is well preserved after irradiation (see Supplementary Information of \cite{Bali2014}). The 
low temperature upturn and the variations in
the NR~$(T)$ curvature evidence that the standard metallic behaviour of the alloy is changed, similarly to other concentrated transition metal alloys\cite{Mooij1973, Houghton1970, Ahmad1974}. 
The NR~$(T)$ changes significantly with higher 
irradiation fluences due to higher number of Fe-Fe nearest neighbours. The sample irradiated with 
9~$\cdot$~$10^{15}$~ions/\si{\centi\metre\squared} exhibits NR~$(\SI{10}{\kelvin})$~$>$~1 and in addition a local broad maximum at $T~\approx~\SI{250}{\kelvin}$.
Both the 
residual resistivity and maximum resistivity have smaller magnitude 
for irradiation of 2~$\cdot$~$10^{16}$~ions/\si{\centi\metre\squared} although the differences are very small.
The occurrence of NR~$(T)$ minimum above the base temperature for irradiated samples is attributable to small mini-bands structure close to the Fermi level\cite{Goedsche1975, Brouers1978}. 
As the thermal activation energy decreases, an electron pocket 
close to $E_{\mathrm{F}}$ gets partially depleted and the composition starts 
exhibiting close to semi-metallic behaviour \cite{Wallace1947, McClure1976, Kaiser2009}. 

\section{Spontaneous Hall Angle}
\begin{figure}[htb!]
\includegraphics[width=\columnwidth]{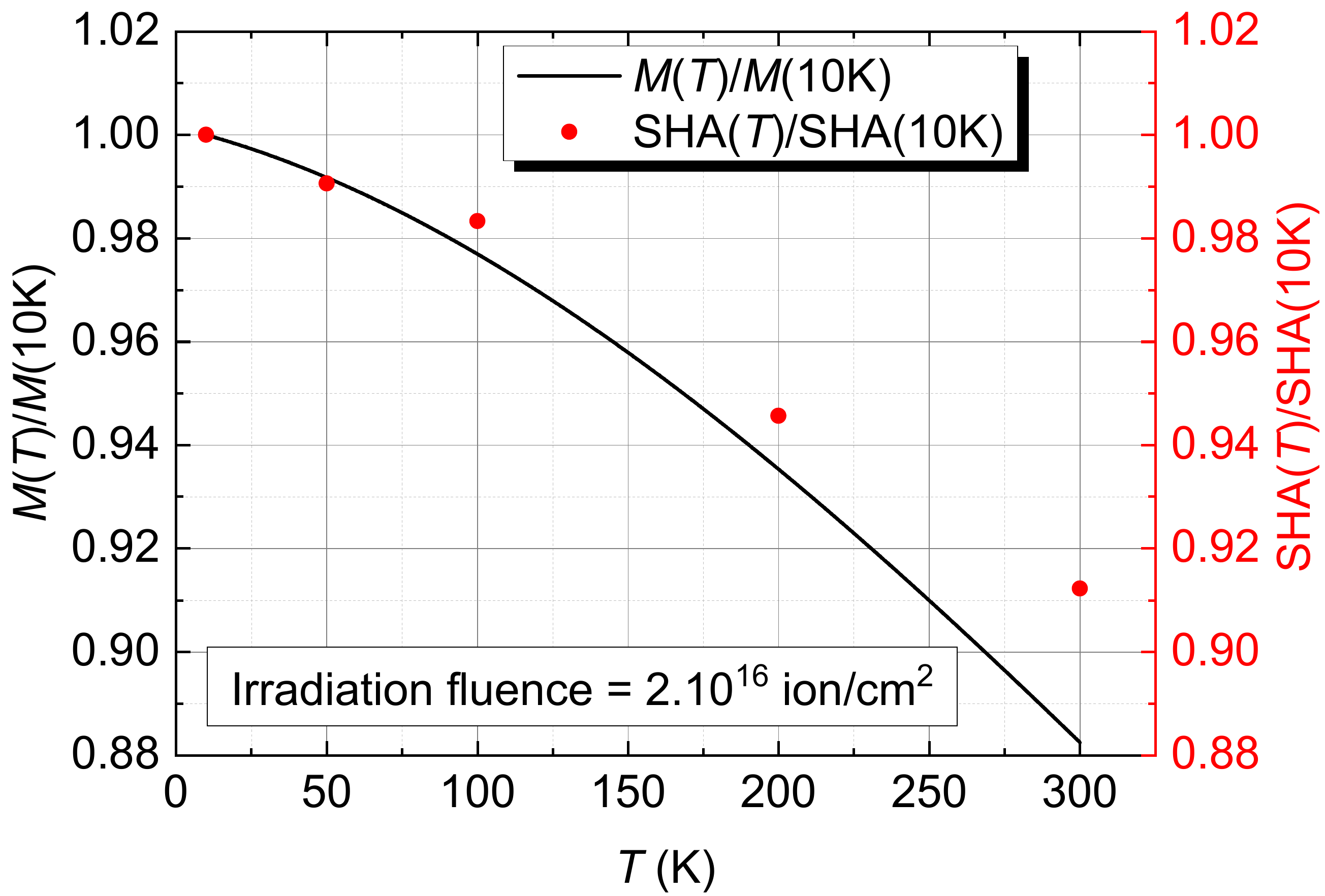}
\caption{Comparison between the temperature evolutions of the magnetization (black line) and SHA (red dots) for irradiation fluence of 2 $\cdot$ $10^{16}$ ions/\si{\centi\metre\squared}. 
}
\label{fig:SHA_SQUID_Suppl}
\end{figure}

The magnetometry temperature evolution and the SHA temperature evolution of the sample irradiated with  2 $\cdot$ $10^{16}$ ions/\si{\centi\metre\squared} are shown in Fig.~\ref{fig:SHA_SQUID_Suppl}. This detailed representation emphasizes the difference between the magnetization and the SHA. The SHA is significantly more field-resilient as it is determined by complicated structure of the density of states close to $E_{\mathrm{F}}$.

\section{M\"{o}ssbauer depth profile spectroscopy}
\begin{figure*}[htb!]
\includegraphics[width=\textwidth]{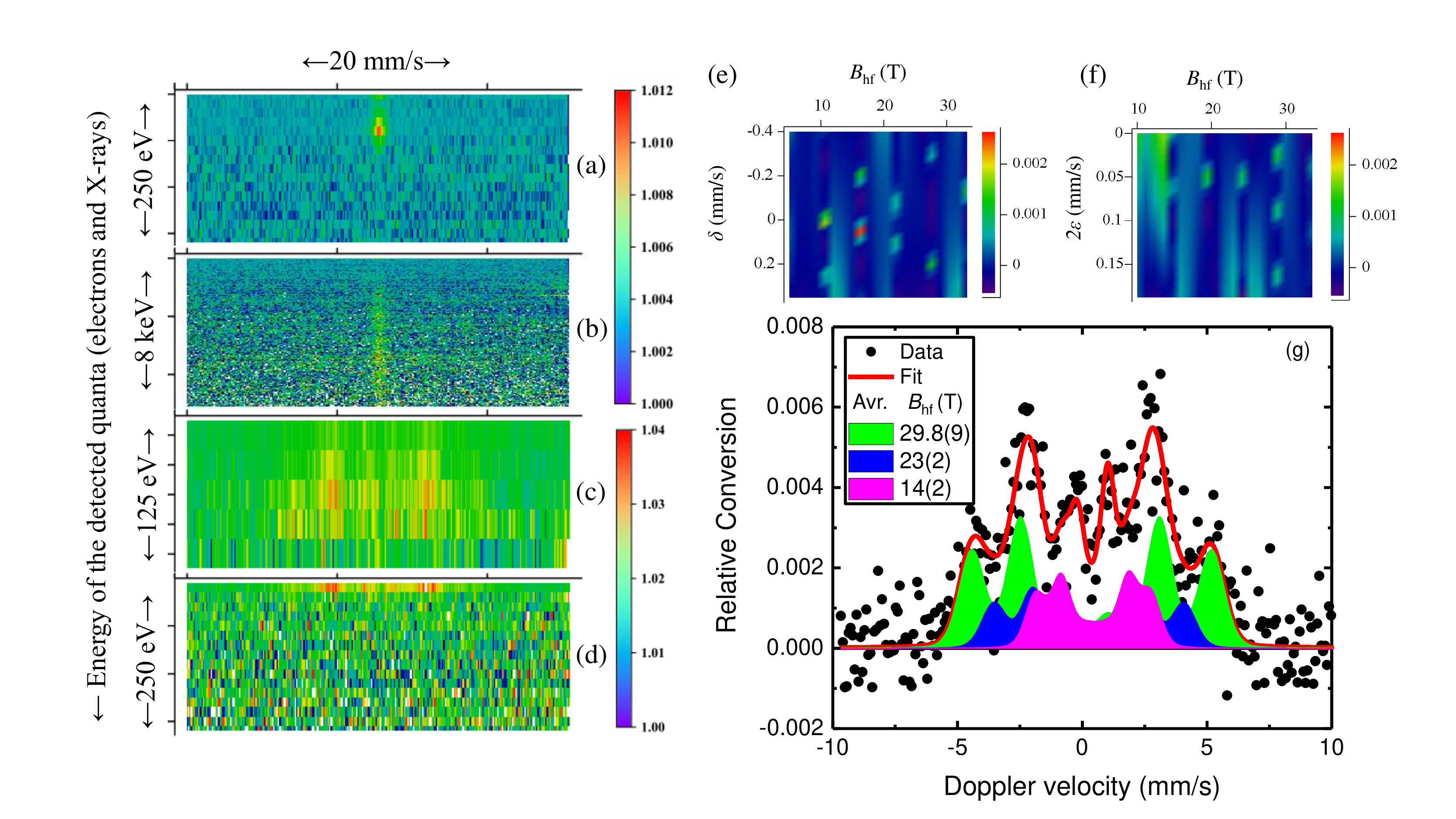}
\caption{
Energy Dispersive Conversion Electron  M\"{o}ssbauer profiles of a non-irradiated (a)
and (b), and irradiated with the highest fluence (c) and (d)
films. The horizontal axis spans the Doppler modulation
velocity and the vertical ones are representing the energy
of the detected electron or X-ray events.
Hyperfine field – isomer shift (e), and hyperfine field –
quadrupole line shift (f) maps for the sample with the
highest irradiation fluence. The colourmap peaks represent
the most probable components of the distribution of
sites. Panel (g) shows a representative
average spectrum for the bulk of the latter film as seen
in electron yield (in black dots). The cumulative fit (red) is the sum of three components: $B_{\mathrm{hf}}$~=~\SI{30}{\tesla} (green), $B_{\mathrm{hf}}$~=~\SI{23}{\tesla} (blue) and $B_{\mathrm{hf}}$~=~\SI{14}{\tesla} (purple).}
\label{fig:FeAl_Mossbauer}
\end{figure*}

Energy Dispersive Conversion Electron  M\"{o}ssbauer Spectroscopy (EDCEMS) has been acquired using a $^{57}$Co source ($\approx \SI{10}{\milli\curie}$) in a Rh matrix at room temperature, in the constant acceleration mode. Both low-energy conversion electrons and X-rays have been detected, using a Wissel gas-filled proportional counter, operated at \SI{1.5}{\bbar} (above atmosphere pressure) of 5$\%$ NH$_3$ in He, flown at 10 sccm. For Fe $K_{\alpha}$, the absorption length within the pressure chamber of the detector would be approximately 9.7 mm. This allows for the absorption and detection of about 10 \% of the X-rays produced.
At this pressure, the absorption depth for low-energy X-rays is sufficiently short to enable their detection with an appreciable sensitivity. This is unlike the common case of operating proportional counters close to or even below atmospheric pressure, when essentially only electrons are detected. The energy resolution for the detection of electrons is rather poor when compared to what is achievable using cylindrical or hemispherical analyzers and does not allow for distinction of the different families of Auger and conversion electrons. There is no counter electric field used for additional energy selection withing the detector. Limited depth information can be recovered by analyzing separately the low and high energy electrons and comparing the resulting spectra with the ones due to detected characteristic X-rays (predominantly Fe $K_{\alpha}$). The spectra measured in X-ray yield sample the entire film thickness essentially uniformly. A purpose-built multi-parameter analyzer has been used to acquire simultaneously the Doppler velocities and electron-escape/X-ray energies, for each and every detector event. The raw pre-amplified pulses are processed through Lower Level and Upper Level Discriminators (LLD, ULD) at the front end of the pulse-sampling ADCs. These are set to just above the pre-amp noise level and at the maximal acceptable pulse height, and do not influence the acquired range of pulse heights. The digital acquisition system is comprised of a number of ADC units for sampling the peak pulse amplitudes or peak areas, as required, digital counter/scalers for synchronisation with the Doppler drive system, a set of 8-bit digital latches, which are latched on the data formation transition of the leading ADC, and a corresponding set of digital access ECP/PCI parallel inputs, up to 5 for a standard desktop PC. Upon the formation of a trigger event, between 3 and 5 digital bytes are latched and recorded into buffer memory on the host PC, under direct interrupt control. A parallel high-level acquisition routine allows for the saving of the raw events as time-stamped log-entries in a raw data file. This data is then binned, as required into 2- or more dimensional arrays, for post-processing. In a typical raw data event, 10-12 bits are used for encoding the energy, 9-12 for encoding the Doppler velocity. These are processed as 3-byte long data word. Two additional bytes are available for extra information, such as the output of time-to pulse-height converters, for example. This allows for the formation of two-dimensional velocity/energy histograms and the rough depth profiling of the films, courtesy of the much smaller escape depths of the low-energy conversion electrons.

We investigate the depth profile of the magnetization and the hyperfine field distribution of the 
irradiated samples by EDCEMS. Conventional CEMS has been previously measured on similarly prepared samples\cite{Liedke2015}, however, the advantages of our purpose-built EDCEMS setup are utilized here for gaining 
further insights.
Figure \ref{fig:FeAl_Mossbauer} (a)-(d) shows the M\"{o}ssbauer detected quanta energy-dispersive spectra at energies dominated by either electrons (sensitive predominantly 
to the top part of the films for detected energies $E~\lesssim~\SI{6}{\kilo\electronvolt}$) and by characteristic X-rays (the entire depth of the films detected energies $E~\gtrsim~\SI{6}{\kilo\electronvolt}$) for both the 
non-irradiated and heavily irradiated films. The entire depth of the non-irradiated film is paramagnetic, with a 
rather low Curie temperature, as is corroborated by an extremely tight doublet seen both in electrons and X-ray 
yield panels (a) and (b). The irradiated film shows a clear non-zero hyperfine field and at least one sextet in electron yield (panel c). The lower the escape energy (electrons coming from deeper down below the surface), the better the spectral definition. In 
X-rays (at higher energies), the contrast is lost as of the spreading of the absorption into the different 
constituent sextets (panel d).
The maps of hyperfine field and isomer shift or quadrupole line shift (panel (e) and (f)) reveal a number of local nearest-neighbour environments ($B_{\mathrm{hf}}$(\si{\tesla}), $2\epsilon$(\si{\milli\meter\per\second}), $\delta$(\si{\milli\meter\per\second})) = (16, 0.05, 0.05); (20, 0.05, 0.1); (25, 0.05, -0.2); (28, 0.025, -0.2), where $B_{\mathrm{hf}}$, $2\epsilon$, $\delta$ are the hyperfine field, quadrupole line shift and the isomer shift relative to bcc-Fe at \SI{300}{\kelvin}. The fitting errors are $\Delta B_{\mathrm{hf}}$~$\approx$~\SI{1}{\tesla}, $\Delta 2\epsilon$~$\approx$~\SI{0.02}{\milli\meter\per\second} and $\Delta \delta$~$\approx$~\SI{0.04}{\milli\meter\per\second}. Apart from the very top surface layers of the film and the interface to the substrate, the hyperfine field is well developed and growing with increasing electron escape energy. 
The mapping clearly demonstrates that there are several different local environments present in the main Fe$_{60}$Al$_{40}$ phase. Moreover, the 
average, spectrum shown on panel (g) is consistent with the majority of the iron moments pointing in-plane\cite{LaTorre2018}, 
even in samples that otherwise demonstrate very good $A$2 phase formation. 
While the three main components in the explicit fit, illustrated on panel (g), exhibit hyperfine parameters, which are matching within the uncertainties the ones found in the maps above (panels (e) and (f)), the mapping resolves one additional component with a $B_{\mathrm{hf}}~\approx~\SI{20}{\tesla}$. The latter might be an artifact of the mapping process, as it is not required for the interpretation of the spectrum in explicit fitting.
The coexistence of paramagnetic and 
ferromagnetic order has already been suggested for the $B$2 phase\cite{Cser1967}. The 
predominant environment shows $B_{\mathrm{hf}} \approx \SI{15}{\tesla}$, which agrees well with what could be 
estimated based on Curie temperature of $\approx$ \SI{600}{\kelvin}, within a mean field approximation.

\bibliography{FeAl_PRB_Suppl}

\makeatletter\@input{FeAl_main_aux.tex}\makeatother